\documentclass[aps,pre,twocolumn,showpacs,superscriptaddress]{revtex4-1}
\usepackage{hyperref}
\usepackage{graphicx}

\begin{document}
\title{Geometric phase and \textit{o}-mode blue shift \\
in a chiral anisotropic medium inside a Fabry--P\'{e}rot cavity}

\author{I.V. Timofeev}
\affiliation{Kirensky Institute of Physics, Siberian Branch of the Russian Academy of Sciences, Krasnoyarsk 660036, Russia}
\affiliation{Laboratory for Nonlinear Optics and Spectroscopy, Siberian Federal University, Krasnoyarsk 660041, Russia}
\email{tiv@iph.krasn.ru}
\author{V.A. Gunyakov}
\affiliation{Kirensky Institute of Physics, Siberian Branch of the Russian Academy of Sciences, Krasnoyarsk 660036, Russia}
\author{V.S. Sutormin}
\affiliation{Kirensky Institute of Physics, Siberian Branch of the Russian Academy of Sciences, Krasnoyarsk 660036, Russia}
\author{S.A. Myslivets}
\affiliation{Kirensky Institute of Physics, Siberian Branch of the Russian Academy of Sciences, Krasnoyarsk 660036, Russia}
\affiliation{Institute of Engineering Physics and Radio Electronics, Siberian Federal University, Krasnoyarsk 660041, Russia}
\author{V.G. Arkhipkin}
\affiliation{Kirensky Institute of Physics, Siberian Branch of the Russian Academy of Sciences, Krasnoyarsk 660036, Russia}
\affiliation{Laboratory for Nonlinear Optics and Spectroscopy, Siberian Federal University, Krasnoyarsk 660041, Russia}
\author{S.Ya. Vetrov}
\affiliation{Kirensky Institute of Physics, Siberian Branch of the Russian Academy of Sciences, Krasnoyarsk 660036, Russia}
\affiliation{Institute of Engineering Physics and Radio Electronics, Siberian Federal University, Krasnoyarsk 660041, Russia}
\author{W. Lee}
\affiliation{Institute of Imaging and Biomedical Photonics, College of Photonics, National Chiao Tung University, Guiren Dist., Tainan 71150, Taiwan}
\author{V.Ya. Zyryanov}
\affiliation{Kirensky Institute of Physics, Siberian Branch of the Russian Academy of Sciences, Krasnoyarsk 660036, Russia}

\date{\today}

\begin{abstract}Anomalous spectral shift of transmission peaks is observed in a 
Fabry--P\'{e}rot cavity filled with a chiral anisotropic medium. The 
effective refractive index value resides out of the interval between the 
ordinary and the extraordinary refractive indices. The spectral shift is 
explained by contribution of a geometric phase. The problem is solved 
analytically using the approximate Jones matrix method, numerically using 
the accurate Berreman method and geometrically using the generalized 
Mauguin--Poincar\'{e} rolling cone method. The $o$-mode blue shift is measured 
for a 4-methoxybenzylidene-4'-$n$-butylaniline twisted--nematic layer inside 
the Fabry--P\'{e}rot cavity. The twist is electrically induced due to the 
homeoplanar--twisted configuration transition in an ionic-surfactant-doped 
liquid crystal layer. Experimental evidence confirms the validity of the 
theoretical model.
\\
\\
The text is available both 
\begin{itemize}
\item in English (Timofeev2015en.tex)
\item and in Russian (Timofeev2015ru.tex; 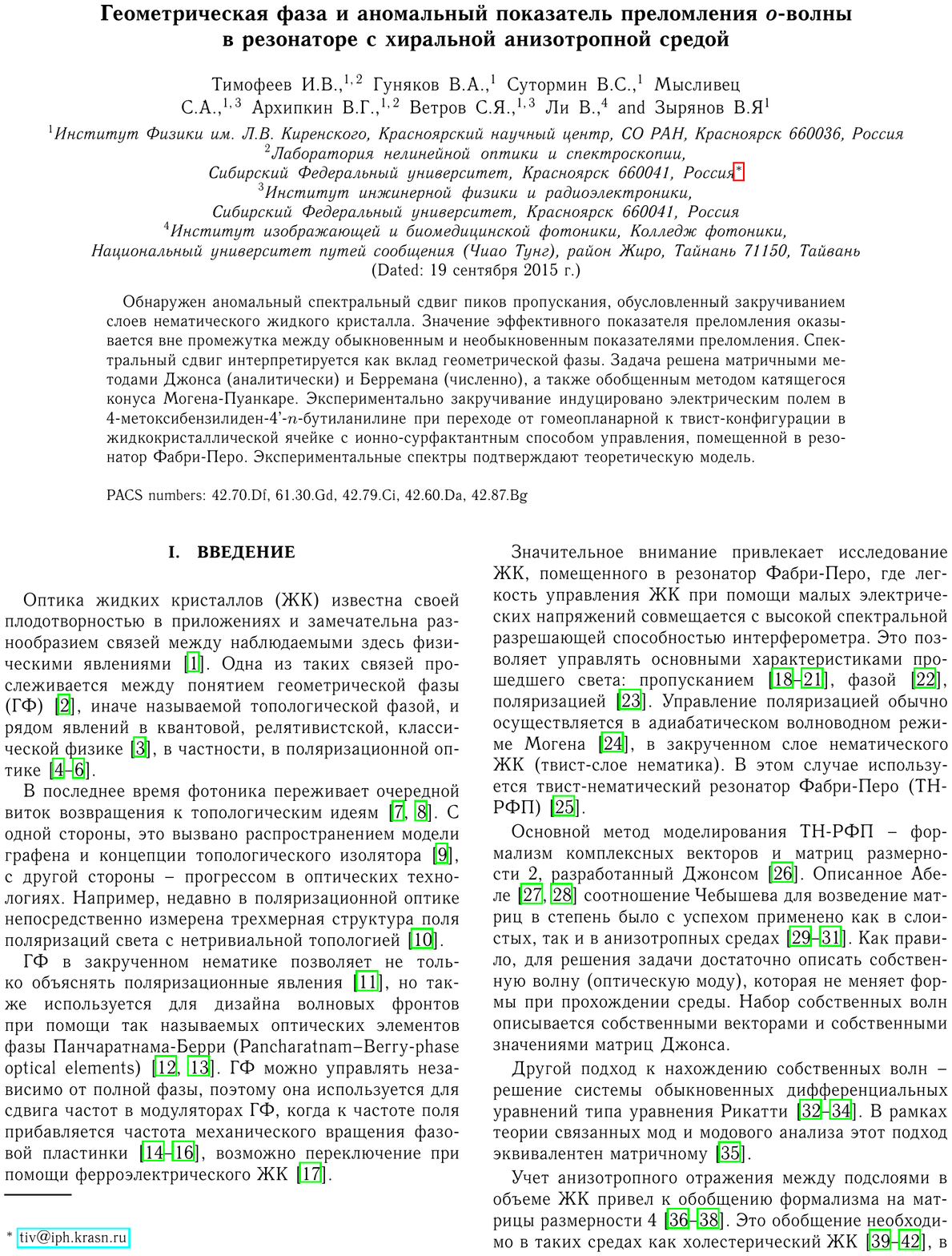)
\end{itemize}
 
\end{abstract}


\maketitle

\section{Introduction}
\label{sec:introduction}
Optics of liquid crystals (LCs) is well known for fruitfulness in 
applications and a remarkable variety of connections between observable 
physical phenomena~\cite{1}. A fascinating connection can be traced between 
the concept of geometric phase (GP)~\cite{2}, 
also known as a topological phase, and a number of phenomena in quantum, 
relativistic, classical physics~\cite{3}, and, in particular, in 
polarization optics~\cite{4,5,6}. Today photonics 
is at the apogee of topological ideas~\cite{7,8}. On the one hand, 
it originates from the condensed-matter graphene idea and the concept of 
topological insulators~\cite{9}. On the other hand, it 
arises from the optical technology advance. For instance, recently in 
polarization optics the three-dimensional structure of the field of light 
polarizations with nontrivial topology has been directly 
measured~\cite{10}.

GP in twisted--nematic polarization phenomena~\cite{11} has 
application for the design of wave fronts using Pancharatnam--Berry phase 
optical elements~\cite{12,13}. 
Remarkably, polarizational GP is independent of the total phase, so it is 
used to offset the frequency of a laser beam by GP modulators, adding the 
mechanical rotation frequency of a quarter--half--quarter-waveplate 
Pancharatnam device~\cite{14,15,16}. It permits 
switching by ferroelectric LC~\cite{17}.

Considerable attention is attracted to the research of LC placed inside a 
Fabry--P\'{e}rot cavity (LC-FPC), combining small-voltage LC manipulation 
and high spectral resolution of the Fabry--P\'{e}rot interferometer. 
Fundamental photonic degrees of freedom---in 
transmittance~\cite{18,19,20,21}, phase~\cite{22}, 
and polarization~\cite{23}---can be efficiently 
controlled. Polarization control usually uses the Mauguin adiabatic 
waveguide mode~\cite{24} in a chiral anisotropic medium, particularly 
in a twisted--nematic (TN) LC layer inside the Fabry--P\'{e}rot cavity 
(TN-FPC)~\cite{25}.

The basic method for TN-FPC calculation is the Jones formalism of complex 
vectors and matrices of dimension~2~\cite{26}. 
Abele~\cite{27,28} had introduced Chebyshev identity for the 
matrix power which was successfully employed both in layered and in 
anisotropic media~\cite{29,30,31}. To solve 
the problem one finds an eigenwave (optical mode) which conserves its shape 
while propagating through the medium. A set of eigenwaves is described by 
the eigenvectors and the eigenvalues of the Jones matrix.

Another approach to find the eigenwaves is to solve the Riccati-like 
ordinary differential equation system~\cite{32,33,34}. Within the 
framework of coupled-mode theory and modal analysis, this approach is 
equivalent to the matrix one (see Appendix C in~\cite{35}).

The account of anisotropic reflections between the sublayers of bulk LC led 
to a generalized Berreman formalism for matrices of dimension 
4~\cite{36,37,84}. This generalization is necessary in 
media such as a cholesteric 
LC~\cite{38,39,40,41} and a TN 
cell of small thickness, and other media with a sharp spatial variation of 
dielectric characteristics at the wavelength 
scale~\cite{42,43,44,Palto2015CLC}. However, Jones formalism gives a good 
approximation when the thickness of the TN cell is several times larger than 
the wavelength and the dielectric constant varies smoothly. 

Assuming no reflection between sublayers of the LC bulk, the TN-FPC behavior 
was described at high voltages~\cite{25} as well as at low 
voltages~\cite{34,45}. The connection between these two extreme 
cases is described in~\cite{46} and generalized in~\cite{47}. 
Another approach is the substitution of the multilayer medium by an 
effective homogeneous anisotropic plate~\cite{48}. The independent 
method is incorporated in~\cite{24,49} using distinct 
mathematical tools: group theory and phase space. Also TN-FPC can be 
considered as a one-dimensional photonic 
crystal~\cite{50,51,52,53}. The 
photonic crystal itself can be formed by a LC material~\cite{54,55}.

The orientation model of the LC layer is required to determine its optical 
response. To the best of our knowledge, in a TN under electrical voltage the 
general orientation model cannot be derived analytically and is simulated 
numerically, even for the one-dimensional 
case~\cite{56}. In contrast to in-plane-oriented 
nematic~\cite{57}, TN produces optical mode coupling manifested 
as avoided crossing of spectral transmission peaks~\cite{56}. A certain way to eliminate the 
mode coupling using anisotropic mirrors is suggested in~\cite{58}. The 
original theoretical study of the apparent paradox of the mode number jump 
for mode coupling inside a TN-FPC is proposed in~\cite{45}.

This article examines the TN-FPC sample with no electric voltage deformation 
that allows analytical description. The direction of the spectral shift of 
the transmission peaks is far from obvious. That's why a visual connection 
is presented for the proposed spectral shift and the geometric phase shift. 
This shift observation is hindered by interplay of four optical waves of 
opposite directions and orthogonal polarizations. Positive feedback 
condition describes the total spectral shift (SS), assuming three types of 
wave coupling. The first type of the coupling originates from LC twist which 
induces a twist spectral shift (TSS). The second coupling type is produced 
by cavity mirror reflection which induces a reflection spectral shift (RSS). 
The third coupling type is produced by reflection between sublayers of LC 
bulk; in this research it is assumed insignificant. Previously developed 
theory~\cite{34,46} is generalized for the account of anisotropy at 
mirror reflection, meaning the distinct reflection phases of $e$- and $o$-wave 
components. The reflection is anisotropic even when the mirror is made of an 
isotropic material while the cavity medium itself is anisotropic.

The experimental scheme excludes significant spectral shift impact from 
parasitic factors other than TSS and RSS. The experiment confirms the theory 
qualitatively and quantitatively.

\section{Analytic model}
\label{sec:analytic}

\begin{figure}[htbp]
\includegraphics[scale=0.50]{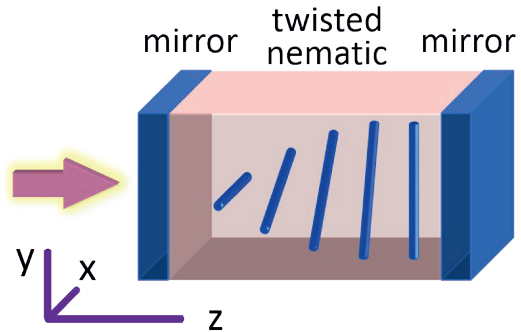}
\caption{Sketch of TN-FPC, a cavity with a chiral anisotropic medium.}
\label{fig1}
\end{figure}

A FPC consists of two plane mirrors (Fig.~\ref{fig1}). The reflecting surfaces are 
oriented in \textit{xy}-plane.
Nematic is placed between the cavity mirrors. The LC director is a unit 
vector of predominant direction of LC molecules. Twist is the state when 
nematic layer is divided into thin lamellar sublayers with the nematic 
director being constant inside every sublayer and rotating from sublayer to 
sublayer. Uniform twist with no orienting external fields is implied when 
the LC director rotates uniformly in the plane of sublayers along the right 
screw. In Fig.~\ref{fig1} the twist angle is 80 degrees, hence the analysis is valid 
for an arbitrary angle. The LC director field determines the local 
dielectric tensor all over the medium. The extraordinary dielectric 
permittivity axis is collinear to the LC director. Consider the nematic with 
a positive uniaxial anisotropy. The extraordinary and ordinary refractive 
indices (RI) correspond to waves with slow and fast phase velocities and 
equal to $n_{e,o} =n\pm \delta n$.

Let the average phase \textit{$\sigma $}, anisotropy phase (angle) $\delta $ and twist angle 
$\varphi $ be linear functions of the coordination $z$ along the layer normal 
direction:
\[
\sigma \left( z \right)=nk_0 \,z,
\quad
\delta \left( z \right)=\delta k\,z,
\quad
\varphi \left( z \right)=k_\varphi \,z,
\]
where $k_0 =\omega /c$ is the angular wavenumber, $\delta k=\delta nk_0 $, 
$k_\varphi =\varphi \left( L \right)/L$, $L$ is the nematic layer thickness or 
distance between mirrors. At $z = L$ let function values $\sigma \left( {z=L} 
\right)$, $\delta \left( {z=L} \right)$, $\varphi \left( {z=L} \right)$ be 
written simply as $\sigma $, $\delta $, $\varphi $ without the function 
argument. Let the light impinge to the TN-FPC strictly in $z $-direction. 
Electric field strength is described by the pair of $x$- and $y$-projections:
\[
E_x \left( z \right)\exp \left[ {i\left( {\omega t-\sigma \left( z \right)} 
\right)} \right]+c.c.,
\]
\begin{equation}
\label{eq1}
E_y \left( z \right)\exp \left[ {i\left( {\omega t-\sigma \left( z \right)} 
\right)} \right]+c.c.,
\end{equation}
where $c.c.$ stands for the complex conjugate component. The pair of complex 
amplitudes $E_{x,y} $ are convenient to be written as the Jones 
vector~\cite{26}:
\[
\vec {e}_{xy} \left( z \right)=\left[ {\begin{array}{l}
 E_x \left( z \right) \\ 
 E_y \left( z \right) \\ 
 \end{array}} \right].
\]
The Jones matrix for untwisted nematic at $\varphi =0$ is diagonal when the 
nematic director is collinear to the $x$-axis:
\begin{equation}
\label{eq2}
\hat {\Delta }\left( \delta \right)=\left[ {{\begin{array}{*{20}c}
 {e^{-i\delta }} \hfill & 0 \hfill \\
 0 \hfill & {e^{+i\delta }} \hfill \\
\end{array} }} \right].
\end{equation}
Note that the extraordinary field component along the $x$-axis has negative 
phase shift according to standard notion (\ref{eq1}):
\[
\vec {e}_{xy} \left( L \right)=\hat {\Delta }\vec {e}_{xy} \left( 0 
\right)=\left[ {\begin{array}{l}
 E_x \left( 0 \right)e^{-i\delta } \\ 
 E_y \left( 0 \right)e^{+i\delta } \\ 
 \end{array}} \right].
\]
The Jones matrix $\hat {\Delta }$ is the transfer matrix (or propagation 
matrix) of the layer. It transfers a polarization state from one layer 
boundary to another by multiplying the corresponding Jones vector.

\subsection{Traveling eigenwaves for a uniformly twisted nematic}
\label{subsec:traveling}
For convenience of subsequent interpretation let us present a direct 
trigonometric derivation of some general expressions for a traveling wave in 
a chiral anisotropic medium. These classical results can be verified 
in~\cite{31}. Let us divide a TN layer into a series of equal sublayers each 
with the thickness \textit{dz} and the anisotropy angle $d\delta =\delta \left( {dz} 
\right)=\delta k\,dz$. Generally, the following is valid for finite not 
twisted sublayers. The twist angle $d\varphi =\varphi \left( {dz} 
\right)=k_\varphi \,dz$ is the angle between the dielectric permittivity 
main axes of neighboring sublayers. Italicized form ``$d$'' is used to 
distinguish it from the particular case of infinitesimal differential 
operator.

The rotation matrix is written as
\begin{equation}
\label{eq3}
\hat {\Phi }\left( \varphi \right)=\left[ {{\begin{array}{*{20}c}
 {\cos \varphi } \hfill & {\sin \varphi } \hfill \\
 {-\sin \varphi } \hfill & {\cos \varphi } \hfill \\
\end{array} }} \right].
\end{equation}
It rotates the reference frame about the $z$-axis by the $\varphi $ angle. The 
Jones matrix for the waveplate (retarder) situated at $\varphi $ angle is 
written as:
\[
\hat {\Delta }_\varphi =\hat {\Phi }^{-1}\hat {\Delta }\hat {\Phi }.
\]
In rotating frame the polarization ellipse appears to be rotated by negative 
angle $-\varphi $. The total Jones matrix is written as the matrix product:
\begin{eqnarray*}
\hat {J}_0 &=&\hat {\Phi }\left( {-\varphi +\frac{d\varphi }{2}} \right)\hat 
{\Delta }\left( {d\delta } \right)...
\nonumber \\
&\times &
\hat {\Phi }\left( {\frac{3d\varphi }{2}} \right)
\hat {\Delta }\left( {d\delta } \right)
\hat {\Phi }\left( {-\frac{3d\varphi }{2}} \right)
\nonumber \\
&\times &
\hat {\Phi }\left( {-\frac{d\varphi }{2}} \right)
\hat {\Delta }\left( {d\delta } \right)
\hat {\Phi }\left( {\frac{d\varphi }{2}} \right)=\hat {\Phi }\left( {-\varphi } \right)\hat 
{J}.
\end{eqnarray*}
%
The product is supposed to be read from right to left with increase of $z$, 
because the Jones column vector is substituted on the right side of the 
matrix. In rotating frame of the matrix $\hat {J}$ the LC director is always 
collinear to the primary axis. It is the $e-o$-frame which is often used in 
description of chiral media~\cite{34,59}.

$z$-axis rotations are additive,
\[
\hat {\Phi }\left( {\varphi _2 } \right)\hat {\Phi }\left( {\varphi _1 } 
\right)=\hat {\Phi }\left( {\varphi _2 +\varphi _1 } \right),
\]
so the Jones matrix is naturally decomposed into the product of certain 
sublayer matrices:
\begin{eqnarray}
\label{eq4}
\hat {J}=d\hat {J}^{N_S }=\left[ {\hat {\Phi }\left( {\frac{d\varphi }{2}} 
\right)\hat {\Delta }\left( {d\delta } \right)\hat {\Phi }\left( 
{\frac{d\varphi }{2}} \right)} \right]^{N_S },
\end{eqnarray}
where $N_S =\varphi /d\varphi $ is the number of sublayers. Substitution of 
Eqs. (\ref{eq2}) and (\ref{eq3}) gives:
\begin{widetext}
\begin{eqnarray*}
d\hat {J}=\left[ {{\begin{array}{*{20}c}
 {\cos \left( {d\varphi /2} \right)} \hfill & {\sin \left( {d\varphi /2} 
\right)} \hfill \\
 {-\sin \left( {d\varphi /2} \right)} \hfill & {\cos \left( {d\varphi /2} 
\right)} \hfill \\
\end{array} }} \right]\left[ {{\begin{array}{*{20}c}
 {e^{-id\delta }} \hfill & 0 \hfill \\
 0 \hfill & {e^{+id\delta }} \hfill \\
\end{array} }} \right]\left[ {{\begin{array}{*{20}c}
 {\cos \left( {d\varphi /2} \right)} \hfill & {\sin \left( {d\varphi /2} 
\right)} \hfill \\
 {-\sin \left( {d\varphi /2} \right)} \hfill & {\cos \left( {d\varphi /2} 
\right)} \hfill \\
\end{array} }} \right],
\end{eqnarray*}
\begin{eqnarray}
\label{eq5}
d\hat {J}=\left[ {{\begin{array}{*{20}c}
 {\cos \left( {d\varphi } \right)\cos \left( {d\delta } \right)-i\sin \left( 
{d\delta } \right)} \hfill & {\sin \left( {d\varphi } \right)\cos \left( 
{d\delta } \right)} \hfill \\
 {-\sin \left( {d\varphi } \right)\cos \left( {d\delta } \right)} \hfill & 
{\cos \left( {d\varphi } \right)\cos \left( {d\delta } \right)-i\sin \left( 
{d\delta } \right)} \hfill \\
\end{array} }} \right].
\end{eqnarray}
\end{widetext}

Eigenvectors $\vec {e}_J $ of the matrix $d\hat {J}$ describe polarization 
conserved in the rotating basis, and eigenvalues $g_J $ of the matrix $d\hat 
{J}$ are phase factors of propagation through the medium layer. The 
eigenvalue condition is:
\begin{eqnarray}
d\hat {J}\;\vec {e}_J =g_J \vec {e}_J ,
\nonumber \\
\det \left( {d\hat {J}-g_J \hat {I}} \right)=0,
\nonumber \\
\det \left( {d\hat {J}} \right)-tr\left( {d\hat {J}} \right)g_J +g_J^2 =0,
\label{eq6}
\end{eqnarray}
where $\hat {I}$ is the identity matrix. According to Eq. (\ref{eq5}) the 
determinant of the matrix equals
\[
\det \left( {d\hat {J}} \right)=1.
\]
Transfer matrix is unimodular when the transferred energy is conserved. The 
trace of the matrix is
\[
tr\left( {d\hat {J}} \right)=2\cos \left( {d\varphi } \right)\cos \left( 
{d\delta } \right).
\]
Let's introduce a new \textit{twisted anisotropy angle} defined as the following:
\begin{equation}
\label{eq7}
\cos \left( {d\upsilon } \right)\equiv \cos \left( {d\varphi } \right)\cos 
\left( {d\delta } \right).
\end{equation}
The solution of Eq. (\ref{eq6}) can then be written as
\[
g_J^\mp =\cos \left( {d\upsilon } \right)\mp i\sin \left( {d\upsilon } 
\right)=\exp \left( {\mp id\upsilon } \right).
\]
The eigenvectors take the forms:
\begin{equation}
\label{eq8}
\vec {e}_J^- =\vec {e}_{te} =\left[ {\begin{array}{c}
 \cos \vartheta \\ 
 -i\sin \vartheta \\ 
 \end{array}} \right],
\quad
\vec {e}_J^\dagger =\vec {e}_{to} =\left[ {\begin{array}{c}
 -i\sin \vartheta \\ 
 \cos \vartheta \\ 
 \end{array}} \right],
\end{equation}
where
\begin{eqnarray}
\label{eq9}
\vartheta &=&\Theta /2,
\nonumber \\
\cos \Theta & \equiv & \sin \left( {d\delta } \right)/\sin \left( {d\upsilon } 
\right),
\nonumber \\
\sin \Theta & \equiv & \sin \left( {d\varphi } \right)\cos \left( {d\delta } 
\right)/\sin \left( {d\upsilon } \right).
\end{eqnarray}

In the literature the pair of eigenwaves described by~(\ref{eq8}) has several names~\cite{31,34,60,61,62}.
As usual the terminology is chosen depending on domination of the anizotropy angle or the twist angle.
With zero twist $d\varphi =0$ and $\vartheta =0$, the eigenwave $\vec 
{e}_J^- $ is simplified into extraordinary $e$-wave, and $\vec {e}_J^\dagger $ is 
simplified into ordinary $o$-wave. They are named as \textit{quasi-e}- and \textit{quasi-o}-waves~\cite{34},
or \textit{te}- and \textit{to}-waves (``twisted waves'')~\cite{31}.
The latter italicized form is appropriate here.
However elliptically polarized \textit{te}-wave may be
confused with linearly polarized TE-wave (transverse electric mode), whose 
electric field is perpendicular to the reference plane or axis.
Also this abbreviation (extraordinary/ordinary) may be confused with the parity 
abbreviation (even/odd).
The chirality of \textit{te}-wave is opposite to LC director chirality.
The chirality of \textit{to}-wave is the same as LC director chirality.
Consequently, the wave pair use to be termed as ``opposite chirality'' wave and ``same chirality'' wave~\cite{60}.
In cholesteric LC \textit{to}-wave demonstrates the Bragg 
reflection, while\textit{ te}-wave is non-diffractive.
That is another way to distinguish the eigenwaves~\cite{61,62}.

The matrix $d\hat {J}$ is diagonalizable using the unitary matrix $\hat {U}$ 
to transform basic vectors $e_{e,o} $ into $e_{te,to} $:
\begin{eqnarray*}
\hat {U}^{-1}=\left[ {\vec {e}_{te} \;\vec {e}_{to} } \right]=\left[ 
{{\begin{array}{*{20}c}
 {\begin{array}{c}
 \cos \vartheta \\ 
 -i\sin \vartheta \\ 
 \end{array}} \hfill & {\begin{array}{c}
 -i\sin \vartheta \\ 
 \cos \vartheta \\ 
 \end{array}} \hfill \\
\end{array} }} \right],
\nonumber \\
\hat {U}=\left( {\hat {U}^{-1}} \right)^\dagger=\left[ {{\begin{array}{*{20}c}
 {\vec {e}_{te}^\dagger } \hfill \\
 {\vec {e}_{to}^\dagger } \hfill \\
\end{array} }} \right]=\left[ {{\begin{array}{*{20}c}
 {\begin{array}{c}
 \cos \vartheta \\ 
 i\sin \vartheta \\ 
 \end{array}} \hfill & {\begin{array}{c}
 i\sin \vartheta \\ 
 \cos \vartheta \\ 
 \end{array}} \hfill \\
\end{array} }} \right],
\end{eqnarray*}
where the symbol ``$\dagger$'' indicates the Hermitian conjugation.
\begin{equation}
\label{eq10}
d\hat {J}=\hat {U}^{-1}\left[ {{\begin{array}{*{20}c}
 {e^{-id\upsilon }} \hfill & 0 \hfill \\
 0 \hfill & {e^{+id\upsilon }} \hfill \\
\end{array} }} \right]\hat {U}.
\end{equation}
The diagonalization simplifies the Jones matrix (Eq. (\ref{eq4})) exponentiation:
\begin{widetext}
\[
\hat {J}=\left\{ {d\hat {J}} \right\}^{N_S }=\left\{ {\hat {U}^{-1}\left[ 
{{\begin{array}{*{20}c}
 {e^{-id\upsilon }} \hfill & 0 \hfill \\
 0 \hfill & {e^{+id\upsilon }} \hfill \\
\end{array} }} \right]\hat {U}} \right\}^{N_S }=\hat {U}^{-1}\left[ 
{{\begin{array}{*{20}c}
 {e^{-id\upsilon }} \hfill & 0 \hfill \\
 0 \hfill & {e^{+id\upsilon }} \hfill \\
\end{array} }} \right]^{N_S }\hat {U}=\hat {U}^{-1}\left[ 
{{\begin{array}{*{20}c}
 {e^{-i\upsilon }} \hfill & 0 \hfill \\
 0 \hfill & {e^{+i\upsilon }} \hfill \\
\end{array} }} \right]\hat {U},
\]
\[
\hat {J}=\left[ {{\begin{array}{*{20}c}
 {\begin{array}{c}
 \cos \vartheta \\ 
 -i\sin \vartheta \\ 
 \end{array}} \hfill & {\begin{array}{c}
 -i\sin \vartheta \\ 
 \cos \vartheta \\ 
 \end{array}} \hfill \\
\end{array} }} \right]\left[ {{\begin{array}{*{20}c}
 {e^{-i\upsilon }} \hfill & 0 \hfill \\
 0 \hfill & {e^{+i\upsilon }} \hfill \\
\end{array} }} \right]\left[ {{\begin{array}{*{20}c}
 {\begin{array}{c}
 \cos \vartheta \\ 
 i\sin \vartheta \\ 
 \end{array}} \hfill & {\begin{array}{c}
 i\sin \vartheta \\ 
 \cos \vartheta \\ 
 \end{array}} \hfill \\
\end{array} }} \right],
\]
\end{widetext}
\begin{equation}
\label{eq11}
\hat {J}=\left[ {{\begin{array}{*{20}c}
 {\cos \upsilon -i\sin \upsilon \cos \Theta } \hfill & {\sin \Theta \sin 
\upsilon } \hfill \\
 {-\sin \Theta \sin \upsilon } \hfill & {\cos \upsilon +i\sin \upsilon \cos 
\Theta } \hfill \\
\end{array} }} \right].
\end{equation}
The uniform twist condition was used:
\[
N_S =\frac{\varphi }{d\varphi }=\frac{\delta }{d\delta }=\frac{\upsilon 
}{d\upsilon }=\frac{z}{dz}.
\]
The solution (\ref{eq11}) is valid for sublayers of finite thickness and is the 
representation of Chebyshev identity~\cite{29}. Equation (\ref{eq7}) for the 
angle $d\upsilon $ is the Pythagorean theorem for the spherical right 
triangle. For the smooth twist function $\varphi \left( z \right)$ the 
sublayer thickness \textit{dz} tends to vanish. The solution can be simplified using 
the Pythagorean theorem for the plane right triangle:
\[
d\upsilon ^2=d\delta ^2+d\varphi ^2.
\]
It is easy to derive it by tailoring the cosines in Eq. (\ref{eq7}). Multiplication 
by $N_S^2 $ produces
\begin{equation}
\label{eq12}
\upsilon ^2=\delta ^2+\varphi ^2.
\end{equation}
The total wave phase then is written as
\begin{equation}
\label{eq13}
\sigma \pm \upsilon =\sigma \pm \sqrt {\delta ^2+\varphi ^2} .
\end{equation}
Let's name it as the Mauguin formula. The effective RI is found by dividing 
both sides by $k_0 L$:
\begin{equation}
\label{eq14}
n_{te,to} =n\pm \sqrt {\delta n^2+\left( {\varphi /k_0 L} \right)^2} .
\end{equation}
The eigenwave ellipticity parameter $\Theta $ from Eq. (\ref{eq9}) can be reduces as 
relation
\begin{equation}
\label{eq15}
\tan \Theta =\varphi /\delta .
\end{equation}
Physically this ellipticity is the smoothness of the twist angle growth in 
comparison with the anisotropy angle growth. It is the adiabatic parameter 
of Mauguin's waveguide regime.

\subsection{Mirror reflection matrix}
\label{subsec:mirror}
The mirror reflection matrix has two multipliers: the phase marix $\hat 
{M}_0 $ and the half-turn rotation matrix $\hat {R}$.
For a mirror made of metal with RI $n_m $ the reflection originates from 
high RI contrast with LC ($n_m -n \gg \delta n)$. In this case the phase matrix 
$\hat {M}_0 $ is approximately isotropic. However for the dielectric 
multilayer Bragg mirror, the RI contrast is not high and the phase matrix is 
far from isotropic. 
\begin{eqnarray*}
\hat {M}_0 &=&\left[ {{\begin{array}{*{20}c}
 {\exp \left( {-i\mu _e } \right)} \hfill & 0 \hfill \\
 0 \hfill & {\exp \left( {-i\mu _o } \right)} \hfill \\
\end{array} }} \right]\\
&=&\exp \left( {-i\sigma _\mu } \right)\left[ 
{{\begin{array}{*{20}c}
 {\exp \left( {-i\delta _\mu } \right)} \hfill & 0 \hfill \\
 0 \hfill & {\exp \left( {+i\delta _\mu } \right)} \hfill \\
\end{array} }} \right]\\
&=&\exp \left( {-i\sigma _\mu } \right)\hat {M}.
\end{eqnarray*}
Various algorithms are used to find phases $\mu _{e,o} $ for certain 
mirrors~\cite{63,64}.

Assume the half-turn rotation matrix to act on the axis perpendicular to LC 
director.
\[
\hat {R}=\left[ {{\begin{array}{*{20}c}
 {-1} \hfill & 0 \hfill \\
 0 \hfill & 1 \hfill \\
\end{array} }} \right].
\]
Obviously, the double reflection of $\hat {R}^2$ produces the identity 
matrix.

\subsection{Perfect cavity eigenwave}
\label{subsec:perfect}
Assume the cavity is perfect (free of losses). The whole loop of the wave 
propagation through the cavity consists of a couple of passages and a couple 
of reflections. The corresponding matrix is the square of the half-loop 
matrix:
\begin{eqnarray}
\label{eq16_0}
\hat {L}&=&\left\{ {\hat {H}\exp \left( {-i\sigma -i\sigma _\mu } \right)} 
\right\}^2
\nonumber \\
&=&\hat {H}^2\exp \left( {-2i\sigma -2i\sigma _\mu } \right),
\end{eqnarray}
where 
$\hat {H}=\hat {R}\hat {M}\hat {J}$.

If the polarization matrix $\hat {H}$ is not an identity matrix, then its 
eigenvectors coincide with the eigenvectors of $\hat {L}$. The eigenvalues 
of $\hat {L}$ are expressed through the eigenvalues of $\hat {H}$:
\[
g_L =g_H^2 \exp \left( {-2i\sigma -2i\sigma _\mu } \right).
\]
Now one can find the eigenvalues of $\hat {H}$:
\begin{widetext}
\[
\hat {H}=\hat {R}\hat {M}\hat {J}=\left[ {{\begin{array}{*{20}c}
 {-1} \hfill & 0 \hfill \\
 0 \hfill & 1 \hfill \\
\end{array} }} \right]\left[ {{\begin{array}{*{20}c}
 {\exp \left( {-i\delta _\mu } \right)} \hfill & 0 \hfill \\
 0 \hfill & {\exp \left( {+i\delta _\mu } \right)} \hfill \\
\end{array} }} \right]\left[ {{\begin{array}{*{20}c}
 {\cos \upsilon -i\sin \upsilon \cos \Theta } \hfill & {\sin \Theta \sin 
\upsilon } \hfill \\
 {-\sin \Theta \sin \upsilon } \hfill & {\cos \upsilon +i\sin \upsilon \cos 
\Theta } \hfill \\
\end{array} }} \right],
\]
\[
\hat {H}=\left[ {{\begin{array}{*{20}c}
 {-\exp \left( {-i\delta _\mu } \right)\left( {\cos \upsilon -i\sin \upsilon 
\cos \Theta } \right)} \hfill & {-\exp \left( {-i\delta _\mu } \right)\left( 
{\sin \Theta \sin \upsilon } \right)} \hfill \\
 {-\exp \left( {+i\delta _\mu } \right)\left( {\sin \Theta \sin \upsilon } 
\right)} \hfill & {\exp \left( {+i\delta _\mu } \right)\left( {\cos \upsilon 
+i\sin \upsilon \cos \Theta } \right)} \hfill \\
\end{array} }} \right].
\]
\end{widetext}
\[
\det \left( {\hat {H}} \right)=-1,
\]
\[
tr\left( {\hat {H}} \right)=2i\left( {\cos \upsilon \sin \delta _\mu +\sin 
\upsilon \cos \Theta \cos \delta _\mu } \right)\equiv 2i\cos \rho _0 .
\]
The eigenvalues are
\[
g_H =i\left( {\cos \rho _0 \mp i\sin \rho _0 } \right)=\exp \left( {i\left( 
{\pi /2 \mp \rho _0} \right)} \right)\equiv \pm \exp \left( {\pm i\rho } 
\right),
\]
where the \textit{resonator} phase $\rho \equiv \pi/2 - \rho_0$,
\begin{equation}
\label{eq16}
\sin \rho =\cos \rho _0 =\cos \upsilon \sin \delta _\mu +\sin \upsilon \cos 
\Theta \cos \delta _\mu .
\end{equation}
The solution corresponds to a couple of modes \textit{resonating} inside the cavity (standing waves)
and let's denote them as \textit{re}- and \textit{ro}-waves.

In the exotic case where $\delta _\mu =\pi /2$ Eq. (\ref{eq16}) is simplified to 
$\rho _0 = \pm \upsilon $. In~\cite{58} the anisotropic mirrors are suggested 
with phases $\mu _e =\pi ,\;\mu _o =0$; consequently, $\sigma _\mu =\delta 
_\mu =\pi /2$.
\begin{eqnarray}
g_L & = & exp(-2(\sigma +\sigma _\mu \pm \rho))
\nonumber \\
& = & exp(-2(\sigma \mp \rho _0) + \pi(1 \pm 1))
\nonumber \\
& = & exp(-2(\sigma \pm \upsilon)).
\nonumber
\end{eqnarray} 
The reflection coupling is removed and RSS is eliminated.

In the isotropic reflection case of $\delta _\mu =0$ Eq. (\ref{eq16}) reduces to
\begin{equation}
\label{eq17}
\sin \rho =\sin \upsilon \cos \Theta .
\end{equation}
It is equivalent to expressions presented in~\cite{34,46}. 

\subsection{Positive feedback condition}
\label{subsec:positive}
The position of transmission peaks is determined by eigenfrequencies of 
perfect cavity waves (modes). They satisfy the positive feedback 
condition~\cite{65} for the total phase shift to be a multiple of 
2\textit{$\pi $}.
\begin{eqnarray*}
-2\sigma -2\sigma _\mu \mp 2\rho =-2\pi N,
\\
\mp \rho =\sigma +\sigma _\mu -\pi N,
\end{eqnarray*}
where $N$ is the cavity mode number. Using Eq. (\ref{eq16}): 
\begin{equation}
\label{eq18}
\mp \sin \left( {\sigma +\mu -\pi N} \right)=\cos \upsilon \sin \delta _\mu 
+\sin \upsilon \cos \Theta \cos \delta _\mu .
\end{equation}
Here the minus sign corresponds to \textit{te}-wave, whereas the plus sign corresponds 
to \textit{to}-wave. It is possible to solve this trigonometric equation graphically.

\subsection{Dispersion curves and TSS}
\label{subsec:dispersion}
\begin{figure*}[htbp]
\includegraphics{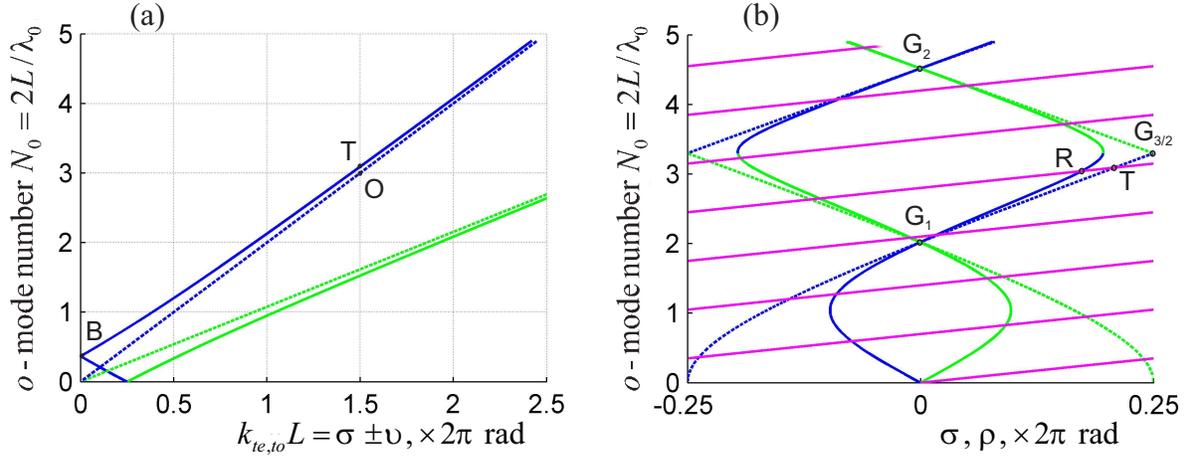}
\caption{The dispersion curves. Abscissa axis shows the phase shift 
proportional to the wave vector. Ordinate axis shows the number of modes, 
proportional to the frequency of the wave. Blue color shows $o$-, \textit{to}-, 
\textit{ro}-waves; green color shows e-, \textit{te}-, \textit{re}-waves.
The calculation parameters are 
$\varphi =\pi /2$, $\delta n/n=0.3$.
(a) TSS calculated with Eq. (\ref{eq13}). 
Points O and T indicate the frequencies of mode 3 for \textit{o}- and \textit{to}-waves, 
respectively. Point B corresponds to the zeroth mode. The splitting at the point B is not shown. 
(b) RSS calculated 
with Eq. (\ref{eq20}). Points T and R show the frequencies of mode 3 for \textit{to}- and 
\textit{ro}-waves, respectively. G$_{1}$ and G$_{2}$ are Gooch--Terry minima (\ref{eq21}) and 
G$_{3/2}$ is the Gooch--Terry maximum (\ref{eq22}).}
\label{fig2}
\end{figure*}

Figure~\ref{fig2}(a) shows dispersion curves of \textit{te}- and \textit{to}-waves for $\varphi =\pi /2$ 
and $\varphi =0$. Scales are dimensionless for both axes. At the ordinate 
the $o$-mode number $N_o =\left( {\sigma -\delta } \right)/2\pi =2L/\lambda _o 
$ is proportional to the frequency of the light field. At the abscissa the 
phase shift is the wave number multiplied by the cavity length.

The branch of \textit{to}-wave does not show the splitting by the cholesteric stopband 
(\cite{1}, at p.~354), for $\varphi =\pi /2$ at $\sigma =\upsilon $, in the 
point B of Fig.~\ref{fig2}(a). The splitting is omitted in Eq. (\ref{eq13}).

Dashed curves for the untwisted structure correspond to \textit{o}- and \textit{e}-waves: 
$\upsilon \left( {\varphi =0} \right)=\delta $. Points O and T indicate 
the TSS of mode 3 from $o$- to \textit{to}-wave. 
Calculated curves for \textit{te}- and \textit{to}-waves lie outside of the intervals of $o$- and 
$e$-wave phases. It illustrates the fact that effective RI lies outside of the 
interval determined by the ordinary and extraordinary RI.

Let's use Eq. (\ref{eq11}) to determine the twist phase shift:
\[
\upsilon -\delta =\sqrt {\delta ^2+\varphi ^2} -\delta ,
\]
assuming $\varphi \ll\delta $ gives
\begin{equation}
\label{eq19}
\upsilon -\delta =\delta \left( {\sqrt {1+\frac{\varphi ^2}{\delta ^2}} -1} 
\right)\approx \delta \left( {\frac{\varphi ^2}{2\delta ^2}} 
\right)=\frac{\varphi ^2}{2\delta }.
\end{equation}

\subsection{Dispersion curves and RSS}
\label{subsec:mylabel2}
Figure~\ref{fig2}(b) illustrates dispersion curves for \textit{re}- and \textit{ro}-waves under 
simplification by Eq. (\ref{eq17}):
\begin{equation}
\label{eq20}
\mp \arcsin \left( {\sin \upsilon \cos \Theta } \right)=\sigma -\pi N.
\end{equation}
The left hand side of Eq. (\ref{eq20}) at $\sigma _\mu =0$ was shown earlier 
in~\cite{34} as a resonance diagram. The right hand site of Eq. (\ref{eq20}) 
produces constant-slope lines. The reduced zone with the period $\pi $ 
corresponds to the half-loop. The positive feedback condition (\ref{eq20}) is 
fulfilled at every intersection of the magenta curve with the green or blue 
one for \textit{re}- and \textit{ro}-wave frequencies corresponding to spectral transmission 
peaks.
Dashed dispersion curves correspond to \textit{te}- and \textit{to}-waves $\mp \arcsin \left( 
{\sin \upsilon \cos \left( {\Theta =0} \right)} \right)=\mp \upsilon $. 
These lines describe the zeroth-order approximation, no reflection coupling.
Points T and R indicate the RSS of mode 3 from \textit{to}- to \textit{ro}-wave.
 
Dispersion curves for \textit{te}-, \textit{to}-, \textit{re}- and \textit{ro}-waves meet at points G$_{1,2\ldots }$ 
where the Gooch--Terry minimum condition is valid:
\begin{equation}
\label{eq21}
\sin \left( \upsilon \right)=0.
\end{equation}
The condition was obtained for minimal TN cell transmittance~\cite{32}. 
In Gooch--Terry minima both resonator phase $\rho $ and twisted anisotropy phase $\upsilon $ 
are multiples of $\pi $. The Jones matrix $\hat {J}$ (Eq. (\ref{eq11})) is 
degenerated into a unit matrix:
\[
\hat {J}=\pm \hat {I}
\]
with $\pm 1$ eigenvalues and arbitrary eigenvectors. Let the condition
\begin{equation}
\label{eq22}
\sin ^2\upsilon =1
\end{equation}
be the Gooch--Terry maximum condition. Note that this simple condition 
describes local maxima of the TN cell transmittance only approximately. In 
fact, the Gooch--Terry transmittance is a sinc function of phase, and maxima 
of this function are slightly non-equidistant. Figure \ref{fig2}(b) shows that in 
Gooch--Terry maxima the phase $\rho $ is maximally distant from the phase 
$\pm \upsilon $. This difference is given by substituting the condition (\ref{eq22}) 
into Eq. (\ref{eq20})
\begin{eqnarray*} 
\upsilon &=&\arcsin \left( 1 \right)=\pi /2,
\nonumber \\
\rho &=&\arcsin \left( {\cos \Theta } \right)=\pi /2-\Theta ,
\nonumber \\
\upsilon -\rho &=&\Theta .
\end{eqnarray*}
The maximum difference is described by the adiabatic parameter $\Theta $. Near the 
Gooch--Terry maximum the difference decreases according to the law:
\begin{equation}
\label{eq23}
\min \left( {\pi -\rho -\upsilon ,\upsilon -\rho } \right)=\frac{\sqrt 
{1+\Theta ^2\tan ^2\upsilon } -1}{\left| {\tan \upsilon } \right|}.
\end{equation}
Near the Gooch--Terry maximal point G$_{3/2}$, the four dispersion curves 
form a typical pattern called avoided crossing. In oscillation theory \textit{re}- and 
\textit{ro}-waves correspond to normal frequencies, while \textit{te}- and \textit{to}-waves correspond to 
partial frequencies~\cite{66}. In quantum mechanics \textit{re}- and \textit{ro}-waves correspond 
to adiabatic states while \textit{te}- and \textit{to}-waves correspond to diabatic 
states~\cite{67}. Both terminologies are used in optics.

The alternation of Gooch--Terry minima and maxima in Fig.~\ref{fig2}(b) can be 
interpreted as an alternation of crossings and avoided crossings of 
transmission peaks~\cite{45,56}. Conditions 
(\ref{eq21}) and (\ref{eq22}) for the untwisted medium correspond to the respective 
conditions of $\sin ^2\delta =0$ and $\sin ^2\delta =1$. The Gooch--Terry 
minimum matches the twisted analogue of a waveplate. This waveplate retardes 
\textit{te}-wave compared to \textit{to}-wave
by an integer number $N_\delta $ of wavelengths
with phase shift $2\pi N_\delta $.
Gooch--Terry maxima correspond to phase shifts of $\left( {2N_\delta +1} \right)\;\pi $.

\subsection{Spectral shifts}
\label{subsec:spectral}
Consider the total spectral shift $\Delta \lambda $ 
of the TN compared with an untwisted counterpart.
Without loss of generality we consider the $o$-wave.
For the $e$-wave one can derive symmetric formulae with the opposite sigh:
$\Delta \lambda _e = -\Delta \lambda _o$

Dimensionless relative shift of the vacuum wavelength $\lambda $ is:
\[
\frac{\Delta \lambda }{\lambda }\approx -\frac{\Delta k}{k_0 -\delta 
k}=-\frac{\rho -\delta }{\sigma -\delta }.
\]
The average phase $\sigma $ can be excluded by use of the following 
relation:
\[
\frac{\lambda }{\sigma -\delta }=\frac{\lambda ^2}{2\pi n_o L}.
\]
And substituting $\rho $ from Eq. (\ref{eq17}) gives:
\begin{equation}
\label{eq24}
\Delta \lambda =-\frac{\lambda ^2}{2\pi n_o L}\left( {\arcsin \left( {\sin 
\upsilon \cos \Theta } \right)-\delta } \right).
\end{equation}
Using approximations (\ref{eq23}) and (\ref{eq19}) one obtains:
\begin{eqnarray}
\label{eq25}
\Delta \lambda &=&\Delta \lambda _{\mathrm{\mathrm{TSS}}} +\Delta \lambda _{\mathrm{RSS}} \nonumber \\
&=&-\frac{\lambda 
^2}{2\pi n_o L}\left( {\frac{\varphi ^2}{2\delta }\mp \frac{\sqrt {1+\Theta 
^2\tan ^2\upsilon } -1}{\tan \upsilon }} \right).
\end{eqnarray}
Far from the Gooch--Terry maximum condition, the second summand $\Delta 
\lambda _{\mathrm{RSS}} $ can be neglected:
\begin{eqnarray}
\label{eq26}
\Delta \lambda _{\mathrm{TSS}} =-\frac{\lambda}{2\pi n_o L} \left( {\frac{\varphi ^2}{2\delta }} \right)
\nonumber \\
=-\frac{\lambda ^2\varphi ^2}{2\pi n_o 2L\;2\pi \delta 
nL/\lambda }=-\frac{\lambda ^3}{2n\delta n}\left( {\frac{\varphi }{2\pi L}} 
\right)^2.
\end{eqnarray}
For the $o$-wave TSS is wavelength-negative $\Delta \lambda _{\mathrm{TSS}} <0$. And 
peaks are shifted in shortwave region; i.e., blue-shifted. However near the 
Gooch--Terry maximum $\tan \left( \upsilon \right)\to \infty $ so that
\[
\Delta \lambda \sim \frac{\varphi ^2}{2\delta }-\frac{\sqrt {1+\Theta ^2\tan 
^2\upsilon } -1}{\left| {\tan \upsilon } \right|}=\frac{\varphi ^2}{2\delta 
}-\Theta \approx \frac{\varphi ^2-2\varphi }{2\delta }.
\]
For example, at $\varphi =\pi /2$,
\[
\Delta \lambda \sim \varphi -2=\frac{\pi -4}{2}<0.
\]
Thus at $\varphi <2$ (in radians) the RSS component
may locally reverse the total shift direction from blue to red.

\section{Interpretation}
\label{sec:interpretation}

\begin{table}[htbp]
\caption{Three levels of anisotropy complexity.}
\begin{tabular}{c c c c}
\hline\hline 
Anisotropic medium& Homogeneous& Twisted& TN-FPC \\
\hline
Eigenwave& 
\textit{o,e}& 
\textit{to,te}& 
\textit{ro,re} \\
\begin{tabular}{@{}c@{}} Eigen\\ polarization \end{tabular}  
&Linear& 
Elliptic& 
\begin{tabular}{@{}c@{}} Linear\\ at boundaries \end{tabular} \\
\hline\hline
\end{tabular}
\label{tab1}
\end{table}

\begin{figure*}[htbp]
\includegraphics{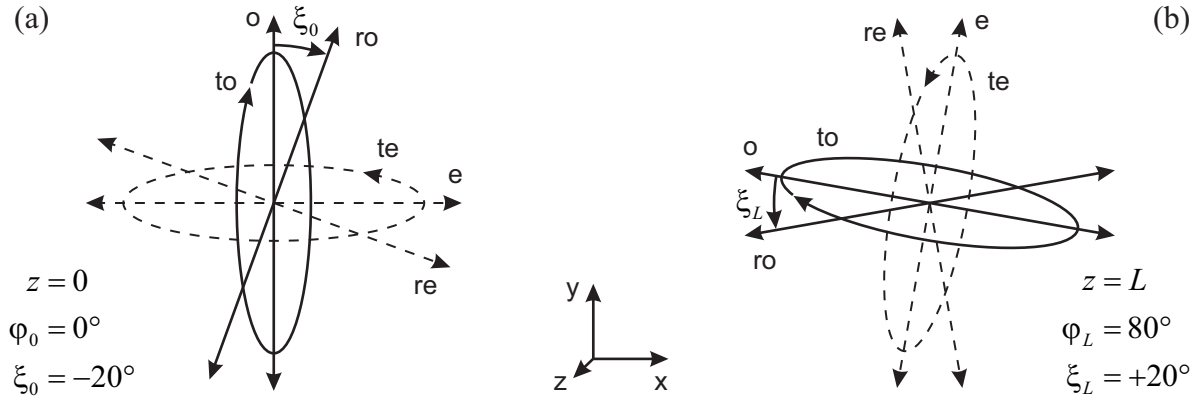}
\caption{Representative polarizations at medium boundaries $z=0$ (a) and 
$z=L$ (b). The twist angle $\varphi $ is between \textit{o}- and \textit{y}-directions.
The \textit{o}-deflection angle $\xi$ is between \textit{o}- and \textit{ro}-directions.
The \textit{e}-deflection angle between \textit{e}- and \textit{re}-directions has the same magnitude as $\xi$ at both boundaries.}
\label{fig3}
\end{figure*}

Three levels of anisotropy complexity are presented in Table~\ref{tab1} and 
illustrated in Fig.~\ref{fig3}.
Eigenwaves \textit{re}- and \textit{ro}- have linear polarizations at mirrors of the 
cavity~\cite{45,46}. These polarizations are biased from the LC 
director and the direction orthogonal to the LC director by the deflection 
angle $\xi $. With increasing frequency linear polarizations rotate 
continuously passing from one of the principal axes to another.
Therefore, we suggest to denote the eigenwaves as $e$2$o$- and $o$2$e$-waves
(English words ``two'' and ``to'' are pronounced identically).
At the Gooch--Terry maximum the polarization 
coincides with the bisectors $\xi =\pm 45^{\circ}$. Therefore, the waves are 
called bisector and orthogonal bisector~\cite{46}. It is convenient to 
treat $e$2$o$-wave as \textit{re}-wave while it is close to $e$-wave and to change its name to 
\textit{ro}-wave after it passes through the bisector and gets close to $o$-wave. And vice 
versa for $o$2$e$-wave. Dispersion curves in Fig.~\ref{fig2}(b) show this renaming by 
changing color at the Gooch--Terry maximum. The curve $\mathrm{G_{1}RG_{2}}$ for 
$o$2$e$-wave is blue-colored in the lower part for \textit{ro}-wave. And it is green-colored 
in the upper part for \textit{re}-wave.

\subsection{Poincar\'{e} sphere}
\label{subsec:poincar}
There is a variety of methods to make the results more comprehensive and 
visually attractive. They are the complex-number representation of 
polarization, Poincar\'{e} sphere~\cite{31,33,68,69}, high-order 
and hybrid-order Poincar\'{e} sphere~\cite{70,71}, admittance diagram 
method, Volpert--Smith chart and three-dimensional (3D) Smith 
chart~\cite{63,72}, and the rolling cone 
method~\cite{24,73}. The last method provides a mechanical 
visualization of Eqs. (\ref{eq12}) and (\ref{eq7}) with a sum of orthogonal angular 
velocities of a solid cone rolling on a plane. For the sublayers of finite 
thickness the cone is replaced by the pyramid. The cone is rolling in a 
characteristic space of light polarization ellipses. This space is usually 
represented as a sphere of unit radius called the Poincar\'{e} sphere (PS).

\begin{figure}[htbp]
\includegraphics{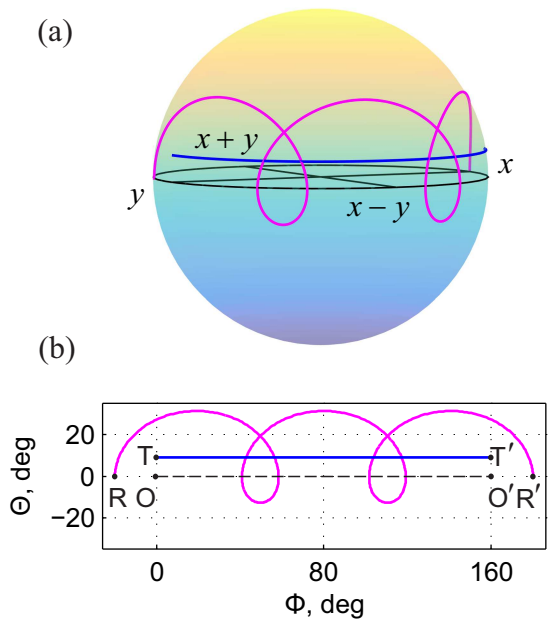}
\caption{Poincar\'{e} sphere (a)
and a portion of its cylindrical projection (b).
Directions $y$, $x - y$, $x$, $x+y$ correspond to angles $\Phi =2\varphi $=0, 90, 
180 and 270 degrees, respectively. Trajectories under the action of operator 
$\hat {J}_0 $ for $\varphi $~=~80$^\circ$: OO$^\prime$ -- linear polarizations on the 
PS ``equator'' corresponding to $o-$wave, TT$^\prime$ -- \textit{to}-wave trajectory on the PS 
``parallel'' with latitude $\Theta $, RR$^\prime$ -- \textit{ro}-wave trajectory, a spherical 
trochoid. The parameters correspond to the rightmost peak in the mentioned 
below experimental spectrum of Fig.~\ref{fig8}(a), $\lambda $~=~579.1 nm.}
\label{fig4}
\end{figure}

Figure~\ref{fig4} shows a \textit{to}-wave polarization trajectory and a \textit{ro}-wave polarization 
trajectory (in blue and magenta, respectively). They smoothly evolve with 
the penetration depth of the TN layer under the action of operator $\hat 
{J}_0 $. For the right-handed \textit{to}- and \textit{ro}-waves it is convenient to set the 
right-handed polarization in the upper hemisphere of PS as 
in~\cite{33,35}, and not in the lower one, as 
in~\cite{6,31,74}. The trajectory 
RR$^\prime$ is the spherical trochoid~\cite{75}. It is associated with a trajectory 
of a point rigidly connected with a solid cone (Fig.~\ref{fig5}(a)) rolling without 
slipping on a plane~\cite{24}. Stereographic projection of a similar 
trajectory is presented in Ref~\cite{33}, p.~136, Fig.~2.24. 

For further consideration it is essential that the unimodular Jones matrix 
has the geometrical sense of the PS rotation. Jones vectors correspond to PS 
points. Distances between points are conserved at transformations on the 
spherical surface. The PS point is often regarded either as the normalized 
triplet of Stokes parameters or as the polarization ellipse traced by the 
terminal point of the field vector. In both cases PS is a two-dimensional 
manifold. But the Jones vector has three degrees of freedom. Its third phase 
is the temporal phase. It progresses by $2\pi $ while the field vector 
passes the elliptic trajectory. The 3-phase polarization state is the 
one-to-one representation of the Jones vector. This polarization state is 
regarded either as the triplet of Euler angles or as the unit quaternion. 
The space of all polarization states is a 3-sphere. The Hopf fibration 
projects it onto PS which is a 2-sphere~\cite{6}. The unit quaternion 
may be imagined as a ``flag'' consisting of two arrows. The first 
polarization arrow goes from the center of PS to its surface. The second 
arrow of temporal phase is connected to the terminal point of the first 
arrow and goes in a perpendicular direction (similar to that shown in Fig.~2 
of Ref~\cite{76}). If the polarization arrow is rotated then 
the entire ``flag'' of polarization state is rotated about the same axis. 
With the temporal phase increase the second arrow rotates around the first 
one. Remarkably, one period on the polarization ellipse corresponds to two 
revolutions of the ``flag.'' Strictly speaking, Jones matrices form the 
special unitary group SU(\ref{eq2}). This group is the universal covering of the 
rotation group SO(\ref{eq3}). The covering is two-sheeted~\cite{77,78,79}.

\subsection{Geometric phase}
\label{subsec:geometric}
The parallel transport of a geometric object on a curved surface rotates the 
object about its own axis. A classic example is the Foucault pendulum with 
the rotation of its swing plane caused by the Earth's daily rotation. 
Similarly, the parallel transport of the polarization state on the PS curved 
surface leads to the phase shift called the geometric phase (GP). It is 
caused by global geometric characteristics, such as the curvature and the 
parallel transport trajectory, and independent of local characteristics, 
such as the speed of state movement along the trajectory. There is a 
\textit{geometric formula} for closed trajectories. Applied to polarization optics it 
claims: GP $\beta$ is equal to minus half the area $\Omega$ encircled by the trajectory on PS:
\begin{equation}
\beta =-\Omega /2.
\label{eqGF}
\end{equation}
A rigorous proof is given in~\cite{6} using Stokes' theorem.

\subsection{Geometric calculation of the phase shift corresponding to TSS for traveling 
wave}
\label{subsec:mylabel3}
We apply the geometric formula to find the phase shift of the traveling wave 
at the trajectory TT$^\prime$ in Fig.~\ref{fig4}. Elliptical wave expressed by Eq. (\ref{eq8}) is a 
superposition of $o-$ and $e-$waves with relevant RIs. The averaged RI should be 
chosen in the form that takes into account the analogue of the 
Aharonov--Anandan dynamic phase (see Ref~\cite{6}, Eq. (5.19)):
\begin{equation}
\label{eq27}
\bar {n}=n-\delta n\cos \Theta ,
\quad
\alpha =\bar {n}k_0 L=\sigma -\delta \cos \Theta .
\end{equation}
The minus sign corresponds to $o$-wave. This RI normalization reduces the PS 
rotation to parallel transport along the great-circle trajectory.

Let the twist angle $\varphi =\pi $. Then $\Phi =2\varphi =2\pi $ makes one 
turn on the PS ``parallel'' with latitude $\Theta $. The area between the 
``equator'' and this ``parallel'' is equal to the side surface of a cylinder 
of unit radius and of height equal to $\sin \Theta $ (see Ref~\cite{80}, 
Lambert cylindrical equal-area projection on p. 260).
\[
\Omega \left( {\varphi =\pi } \right)=2\pi \sin \Theta .
\]
For arbitrary twist angle
\begin{equation}
\label{eq28}
\Omega \left( \varphi \right)=2\varphi \sin \Theta .
\end{equation}
The total phase shift $\gamma $ consists of the dynamic phase $\alpha $ and 
the geometric phase $\beta $: $\gamma =\alpha +\beta $.
The geometric formula (\ref{eqGF}) and equations (\ref{eq27},\ref{eq28}) 
give:
\[
\gamma =\sigma -\delta \cos \Theta -\varphi \sin \Theta .
\]
Eqs. (\ref{eq12}) and (\ref{eq15}) produce a transformation $\delta =\upsilon \cos \Theta $, 
$\varphi =\upsilon \sin \Theta $. This transformation gives
\[
\gamma =\sigma -\upsilon \left( {\cos ^2\Theta +\sin ^2\Theta } 
\right)=\sigma -\sqrt {\delta ^2+\varphi ^2} .
\]
Indeed, the result coincides with the Mauguin formula (\ref{eq13}).

\subsection{Geometric calculation of the phase shift corresponding to RSS for cavity 
wave}
\label{subsec:mylabel4}
\begin{figure}[htbp]
\includegraphics{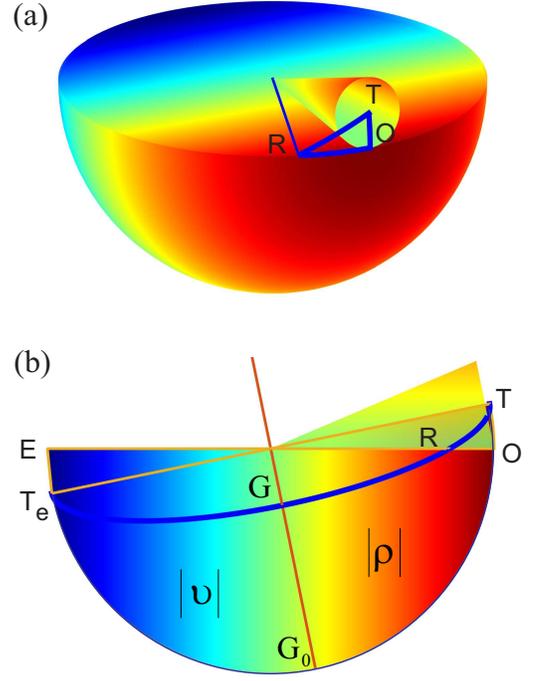}
\caption{ 
The rolling cone method on the Poincar\'{e} sphere. (a) Points O, T 
and R describe the polarizations of $o$-, \textit{to}- and \textit{ro}-waves at the boundary of the 
twisted layer. The Mauguin cone contains the point T on its axis, and the 
point O on its generator. Spherical triangle OTR has a right angle
$\protect\angle $ROT.
The cathetus RO = $\Xi =2\xi $ lies on the ``equator''. The cathetus 
TO = $\Theta =2\vartheta $ is perpendicular to the ``equator''. Acute angles 
are 
$\protect\angle \mathrm{RTO}= \upsilon $ and $\protect\angle \mathrm{ORT}= {\rho _0 } $.
The area is equal to spherical excess
$\Omega \left( \mathrm{OTR} \right)= (\pi/2 + \upsilon + \rho_0) - \pi = \upsilon - \rho $.
(b) Points E and 
T$_e$ describe polarizations of $e$- and \textit{te}-waves. The great circle G$_0$G is 
perpendicular to the diameter TT$_e$. The big circle T$_e$GRT in the 
Gooch--Terry minimum coincides with the great circle T$_e$G$_0$OT. Distance 
to the Gooch--Terry minimum for the \textit{to-}wave is determined by the phase 
$\upsilon =\protect\angle \mathrm{G_0 G}=\Omega \left( {\mathrm{TG_0 G}} \right)$, and for 
\textit{ro}-waves it is determined by the phase $\rho=\Omega(\mathrm{G_0GRO})$.}
\label{fig5}
\end{figure}

The Mauguin--Poincar\'{e} rolling cone method is easily expanded to the Hopf 
bundle of PS. In fact, the generalization is a consequence of the solid cone 
analogy. This generalization allows to find certain phase relations 
trigonometrically. The phase integration along the trajectory RR$^\prime$ 
corresponds to the matrix multiplication of Eq.~(\ref{eq4}). Diagonalization of Eq. 
(\ref{eq10}) is a simplifying algebraic transition to the rotating frame written as 
the product of three matrices. Geometrically, it corresponds to the 
transition from the trajectory RR$^\prime$ to the chain of three arcs: RT-TT$^\prime$-T$^\prime$R$^\prime$ 
(Fig.~\ref{fig4}(b)). Wherein the first and the last arcs are to be chosen as 
geodesics. The great circle arc RT (see Fig.~\ref{fig5}(a)) matches the parallel 
transport and the geometric phase corresponds to the Pancharatnam 
phase~\cite{4} without any normalizing dynamic phase of Eq. (\ref{eq27}).

During the passage of the distance $L$ ahead through the cavity the 
traveling \textit{to}-wave in rotating frame receives the phase shift $\upsilon $. 
Therefore, PS is rotated by the angle $2\upsilon $. After the rotation the 
linear polarization $R$ is returned back to the ``equator'' at the point R$^\prime$. It 
corresponds to reflection symmetry of triangles OTR and O$^\prime$T$^\prime$R$^\prime$. The mirror 
reflection corresponds to the triangle reflection symmetry with respect to 
the RO arc (improper rotation). The resultant geometric phase shift of the 
\textit{ro}-wave compared with that of the \textit{to}-wave accounts the triangle area four times 
per loop cycle $4\Omega \left( {OTR} \right)=2\beta _{\mathrm{RSS}} $.

The rolling cone method works even when the adiabatic condition is violated. 
In~\cite{47} the non-uniform LC twist is examined, for example, under 
electric voltage. However, in the symmetric case for the cone opening angle 
$\Theta \left( z \right)=\Theta \left( {L-z} \right)$ the eigenwave 
preserves linear polarization on the layer boundary.
For practice the linear polarization of the transmitted light is advantageous,
because of the reliable blocking of transmission~\cite{Palto2011}.

The account of mirror phase shift $\delta _\mu $ by anisotropic reflection is 
equivalent to a waveplate action. On PS this leads to an additional rotation 
of the RO arc by the angle of $\delta _\mu $ off the ``equator.'' 
Pythagorean theorem (\ref{eq17}) for the right spherical triangle OTR transforms to 
the cosine theorem (\ref{eq16}) for the angles of the spherical triangle. The point 
R leaves the PS ``equator''. It means that the corresponding transmission 
peak becomes elliptically polarized. The superposition of opposing traveling 
waves on a perfect cavity mirror remains linearly polarized at $\delta _\mu 
\ne 0$ only when $\sigma _\mu =\pi N_\mu/2$ with integer $N_\mu$.

\subsection{Connection between dispersion curve phases and PS angles}
\label{subsec:connection}
With increasing frequency the Mauguin cone rolling angle $2\upsilon $ is 
increased uniformly. The \textit{to}-wave corresponds to the Mauguin cone axis on PS. 
The point T is fixed, assuming constant adiabatic parameter: $\Theta \left( 
\omega \right)\approx \mbox{const}$. Without changing the polarization the 
\textit{to}-wave acquires a $\pi $ phase between the adjacent Gooch--Terry minima (see 
the section G$_{1}$G$_{3/2}$G$_{2}$ of the dispersion curve at Fig.~\ref{fig2}(b)). 
The \textit{ro}-wave gains the phase $\rho $ (dispersion curve section G$_{1}$RG$_{2})$ 
which is less than the phase $\upsilon $ by the area of a spherical triangle 
$\Omega \left( {OTR} \right)= \upsilon - \rho $. 
During the passage of the Gooch--Terry maximum this area has a critical 
value $\Omega \left( {OTR} \right)=\Theta $ in accordance with Eq. (\ref{eq23}).

The cathetus RO=$\Xi =2\xi $ of the triangle OTR is given by Napier's rules 
for right spherical triangles~\cite{79}:
\[
\tan \Xi =-\sin \Theta \tan \upsilon .
\]
This equation determines the deflection angle $\xi $ of the boundary 
\textit{ro}-wave linear polarization from the LC director (Fig.~\ref{fig3}). Figure~\ref{fig4}(b) shows 
the arc TR rotating when $\upsilon $ increases. The point R moves 
non-uniformly along the PS ``equator''. The movement is faster near the 
Gooch--Terry maximum. In adiabatic approximation $\Theta \ll\pi /2$ the point 
R jumps at the Gooch--Terry maximum so that $\xi \approx 0$ for \textit{ro}-wave and 
$\xi \approx \pi /2$ for \textit{re}-wave.

Exact correspondence between optical wave phases and characteristic space 
angles provides a visual support and a qualitative understanding of the 
phenomenon supporting the validity of the obtained solution.

\subsection{Intermediate optical response presumption}
\label{subsec:intermediate}
The motivation for this study was the debate about the direction of the 
transmission peak spectral shift in a TN layer. The \textit{intermediate optical response presumption} was formulated for an 
anisotropic medium whose \textit{effective RI is between the ordinary and extraordinary RI,}
\begin{equation}
\label{eq29}
n_o <\tilde {n}<n_e 
\end{equation}
Here are a few abstract examples supporting the intermediate optical 
response presumption.

\begin{enumerate}
\item The average RI of nematic in the isotropic phase is~\cite{1,73}
\[
n_{iso}^2 =\left( {n_e^2 +2n_o^2 } \right)/3.
\]
\item In a homogeneous, uniaxially anisotropic medium, the extraordinary light 
wave propagating at an angle $\theta $ to the optical axis has the following 
RI (Ref~\cite{1}, Eq. (11.6)):
\[
n_e \left( \theta \right)=n_o n_e \left( 0 \right)/\sqrt {n_e^2 \left( 0 
\right)\cos ^2\theta +n_o^2 \sin ^2\theta } .
\]
\item The thin sublayer of the TN layer has effective RI according to the 
normalization (\ref{eq27}).
\end{enumerate}

Non-additive response of a slab of sublayers leads to the geometric phase 
(\ref{eqGF}). As a result, the Mauguin formula (\ref{eq14}) contradicts the intermediate 
optical response presumption (\ref{eq29}):
\[
n_{te,to} =n\pm \sqrt {\delta n^2+\left( {\varphi /k_0 L} \right)^2} ,
\quad
n_{to} <n_o <n_e <n_{te} .
\]
The stated contradiction admits an experimental test. With an increase in 
the effective RI (i.e., increase in the optical length of the cavity), 
transmission peaks are shifted to the red. For the $o$-wave in untwisted 
structure the effective RI is equal to the ordinary RI. According to the 
intermediate optical response presumption (Eq. (\ref{eq29})) the twisted effective 
RI shifts towards the extraordinary RI; in other words, the effective RI 
increases. This predicts the redshift $\Delta \lambda _{\mathrm{TSS}} >0$ for 
spectral peaks. In contrast, the Mauguin formula predicts the blue shift 
$\Delta \lambda _{\mathrm{TSS}} <0$ (\ref{eq26}).

\section{Experiment}
\label{sec:experiment}
\subsection{LC orientation}
\label{subsec:mylabel1}
\begin{figure*}[htbp]
\includegraphics[scale=0.9]{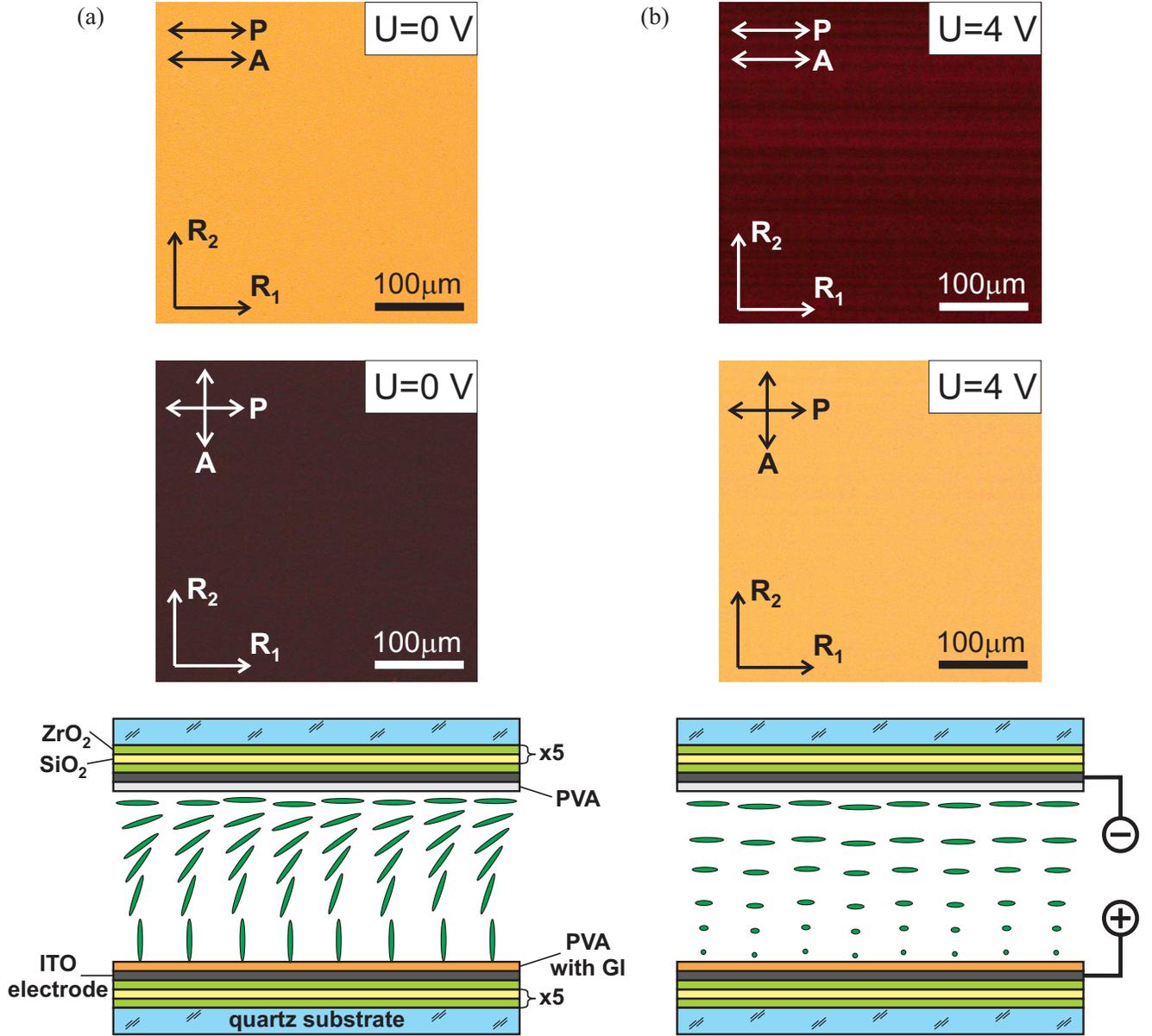}
\caption{Microphotographs of optical textures of LC-FPC under the 
parallel-polarizer scheme (top row) and crossed polarizers (middle row), and 
the respective diagrams of the LC director configurations (bottom row). (a) 
Homeoplanar orientation at U = 0 V; (b) twist orientation at U = 4 V. 
Polarizer (P) and analyzer (A) directions are represented by the double 
arrows. R1 and R2 are rubbing directions of the top and bottom mirrors, 
respectively.}
\label{fig6}
\end{figure*}

An experimental study of the shift of the $o-$polarized spectral transmission 
peaks to shorter wavelengths was carried out inside a LC-FPC. The 
Fabry--P\'{e}rot cavity consisting of two dielectric mirrors (Fig.~\ref{fig6}, bottom 
row) is treated as a LC cell. It was filled with the nematic LC 
4-methoxybenzylidene-4'-$n$-butylaniline (MBBA) doped with the cationic 
surfactant cethyltrimethyl ammonium bromide (CTAB) in the weight ratio 1: 
0.003. The cavity gap is 7 \textit{$\mu $}m. CTAB molecules within MBBA dissociate into 
Br-anions and surface active CTA-cations. The latter are adsorbed at the 
surfaces of aligning layers and, at the sufficient concentration, can form a 
layer specifying the homeotropic coupling condition for nematic 
molecules~\cite{81,82}.

A multilayer mirror coating consists of 6 layers of zirconium dioxide 
(ZrO$_{2})$ with RI of 2.04 and a thickness of 55 nm and 5 layers of silicon 
dioxide (SiO$_{2})$ with RI of 1.45 and a thickness of 102 nm, alternately 
deposited on the surface of a quartz substrate. The alternating layers 
produce the reflection band in the range of 420--610~nm. Thin ($\sim $ 150 
nm) ITO-electrodes were deposited on the upper layers of ZrO$_{2}$ to apply 
the electric field normally to the LC-FPC mirrors. Electrodes were covered 
with different polymeric alignment films by spin-coating to implement the 
initial homeoplanar director orientation (Fig.~\ref{fig6}(a), bottom row). The top 
substrate was covered with a planar-orienting film of pure polyvinyl alcohol 
(PVA). The bottom substrate was covered with a PVA film doped with glycerin 
(Gl) compound in the weight ratio 1: 0.61. With the utilized CTAB 
concentration the surfactant ion molecules were adsorbed on the PVA--Gl 
film. The formed layer shields the planar-orienting effect of the polymer 
coating and provides the homeotropic anchoring for MBBA. Polymer films on 
both dielectric mirrors were rubbed unidirectionally to define the axis of 
easy orientation. The angle between the rubbing directions at the top (R1) 
and bottom (R2) mirrors is 90$^\circ$. 

Inside the LC-FPC cell at constant voltage value U = 4 V, the orientation 
transition from homeoplanar to twist director configuration is induced by 
ionic modification of surface adhesion~\cite{83}. This transition is 
accompanied by a modification in polarizing microscope optical texture of 
the LC-FPC cell under the parallel- and crossed-polarizer scheme (Fig.~\ref{fig6}, 
top row and middle row, respectively). For example, under crossed polarizers 
the optical texture of LC-FPC cells in the initial state is uniformly dark 
when the R1 direction coincides with the transmission axis of the polarizer 
(Fig.~\ref{fig6}(a), middle row). The light transmission is increased when the voltage 
is applied and the twist-configured FPC (TN-FPC) is formed (Fig.~\ref{fig6}(b), middle 
row). However, under the parallel-polarizer scheme the TN-FPC transmittance 
is low (Fig.~\ref{fig6}(b), upper row) due to the rotation of the polarization plane of 
linearly polarized light at an angle close to 90$^\circ$ after passing the TN 
layer.

The experimental scheme excludes a significant impact of parasitic factors 
on the shift. The structure is twisted uniformly, because every sublayer has 
constant twist torque created by different rubbing orientations. 
Twist-structure is under the voltage of 4 V. However, this does not lead to 
any deformation of the structure. Firstly, in the volume of the cell the 
voltage is partially compensated by surface charges. Secondly, MBBA is 
oriented transversely to electric field. 

\subsection{Cavity and its spectra}
\label{subsec:cavity}

\begin{figure}[htbp]
\includegraphics[scale=0.5]{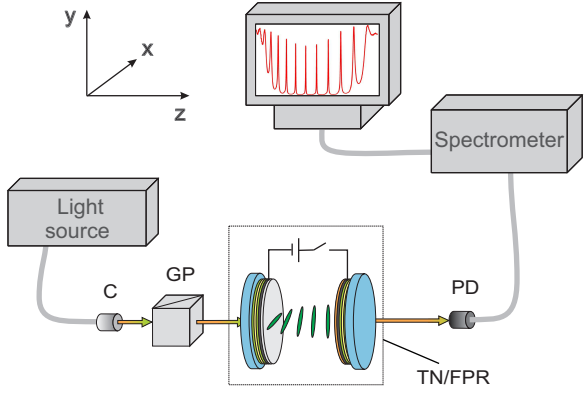}
\caption{Experimental setup for measurement of polarized transmittance 
spectra of LC-FPC. C -- fiber collimator, GP -- Glan prism, PD -- 
photodetector.}
\label{fig7}
\end{figure}

Polarized transmission spectra of LC-FPC with two different configurations 
of MBBA director for normally incident light were measured using an Ocean 
Optics HR4000 spectrometer equipped with fiber-optics (Fig.~\ref{fig7}).
The LC-FPC sample was placed inside the optical channel with planar 
orientant (R$_{1}$~$\vert \vert $~$x)$ at the input mirror. A Glan prism (P) 
was used as a polarizing element with the linear polarization along the 
$y$-axis (i.e., orthogonal to the director on the input mirror). In this 
experimental scheme the $o$-polarized transmission spectrum was measured 
regardless of structural transformations in the nematic volume. Spectra were 
recorded at a fixed temperature of 23.0$^\circ$~C. Thermal stabilization error 
was no more than $\pm $0.2$^\circ$~C. The voltage was generated by a power 
supply (AKTAKOM ATN-1236).

\subsection{Comparison of experiment, simulation and analytic formulae}
\label{subsec:comparison}
\begin{figure}[htbp]
\includegraphics{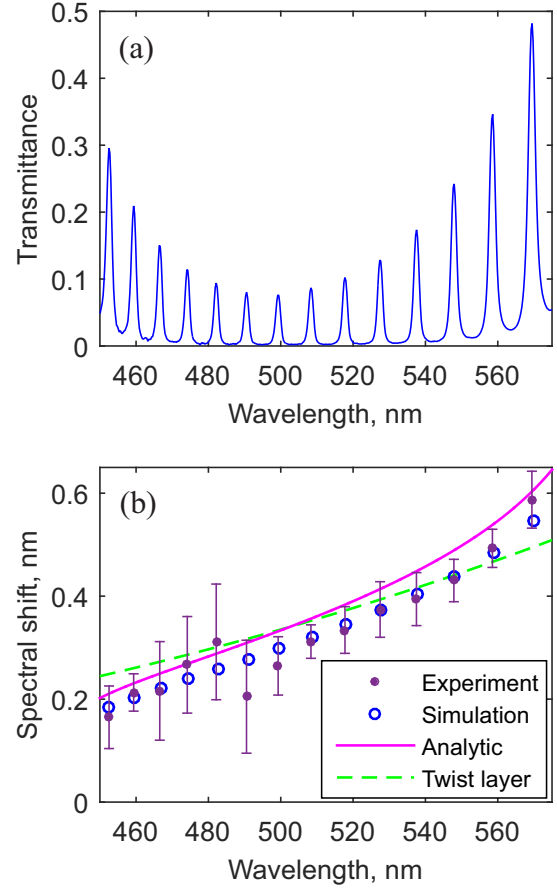}
\caption{Ordinary-polarized LC-FPC spectrum (a)
and the spectral shift of transmission peaks (b).
($\bullet$) -- experimental shift values, (o) - calculated 
shift values obtained by direct numerical simulation using the Berreman 
method, solid curve -- analytical shift calculated from Eq. (\ref{eq24}) obtained by 
the Jones method, dashed curve -- the shift $\Delta \lambda _{\mathrm{TSS}} $ 
calculated without mirror effect from Eq. (\ref{eq26}).
The parameters of the rightmost peak with $\lambda $~=~579.1 nm
were taken to calculate Fig.~\ref{fig4}.}
\label{fig8}
\end{figure}

Three measurements were averaged to calculate the experimental spectral 
shift (Fig.~\ref{fig8}). Experimental resolution of 0.25 nm was improved by fitting 
spectral peaks by Voigt contours. The confidence interval was calculated as 
the corrected sample standard deviation multiplied by the Student 
coefficient with 95{\%} reliability: $t_{0.05,2} $= 4.3027.

The Berreman method~\cite{37,84} was used for 
numerical simulations. The method evaluates polarization vectors and 
transfer matrices of dimension 4 to take into account the reflection between 
sublayers of the LC bulk. The TN layer was divided into 200 sublayers and 
the calculated spectral resolution was 0.01 nm. Some parameters were 
considerably tuned to match the experimental spectra. The thicknesses and 
RIs of amorphous layers constituting the dielectric mirror were taken for 
SiO$_{2}$: 83 nm and 1.45; for ZrO$_{2}$: 66 nm and 2.02; for each ITO 
layer: 117 nm and 1.88858 + 0.022i taking the absorption into account; for 
the substrate RI: 1.45 and for the PVA RI: 1.515; two thicknesses of PVA 
layers: 300 and 600 nm; for MBBA extraordinary RI: 1.737 and ordinary RI: 
1.549, both with RI imaginary part: 0.00078i. The thickness of MBBA layer 
was 7980 nm; the twist angle was 80$^\circ$.

The material dispersion gave some minor changes in the spectra and the 
shift. The most notable changes were the absorption dispersion and the 
change in MBBA layer thickness by 40 nm. Therefore, the spectra illustrated 
in Figs. 8 and 9 were calculated assuming no material dispersion.

\begin{figure}[htbp]
\includegraphics{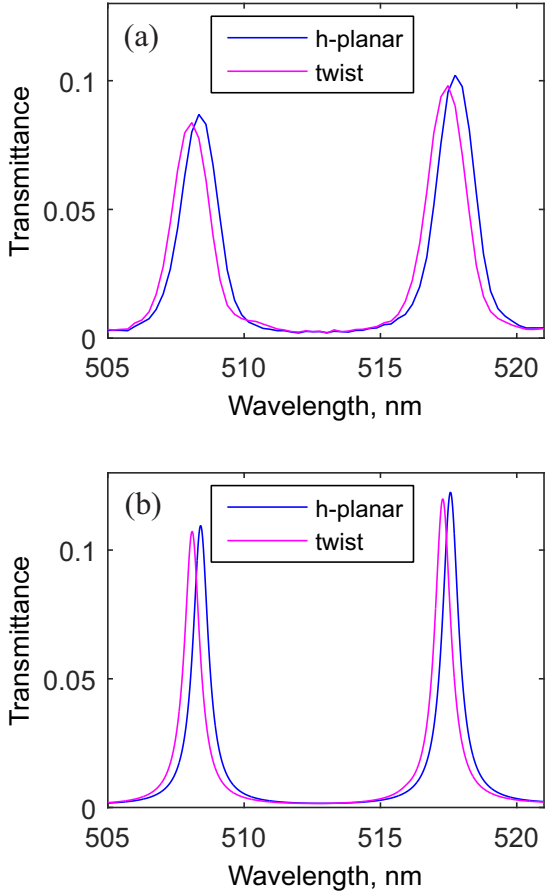}
\caption{Zoomed experimental (a) and calculated (b) transmittance spectral 
peaks for homeoplanar (blue) and twisted (magenta) configurations in LC-FPC. 
The twist leads to shorter wavelength shifts.}
\label{fig9}
\end{figure}

Experimental transmission peaks are broadened no more than 2 times larger 
than the calculated peaks (Fig.~\ref{fig9}). This was achieved by additional flatness 
tuning of the cavity. Comparison of experimental spectra with simulations 
shows the transmission peaks shift to shorter wavelengths for the twisted 
configuration. The shift is less than the half-width at half maximum of a 
peak. Three peaks in the range of 480--500 nm have minimal intensity and 
maximal shift dispersion. However, all 14 points fit satisfactorily into the 
simulated dependence curve.

Figure 8(b) shows by dots and circles the experimental and calculated values 
of the spectral shift, respectively. Analytical Eqs. (\ref{eq24}) and (\ref{eq26}) are shown 
by the solid curve and the dotted curve, respectively. The reflection phase 
shift leads to simultaneous displacement of all the points along the solid 
curve. The latter analytic curve is affected by the reflection phase shift 
only near Gooch--Terry maxima. The experimental spectrum contains no 
Gooch--Terry maxima. So in analytical equations the reflection anisotropy 
was ignored $\delta _\mu =0$.

The analytic formula (\ref{eq24}) slightly overestimates the shift. There are 
several possible reasons for this distinction. Firstly, the difference of 
PVA layer thicknesses on mirrors leads to some error. Secondly, reflection 
between sublayers of LC bulk produces the difference between the Berreman 
and the Jones spectral shifts. The effective Jones RI (\ref{eq14}) is commonly used 
for the TN LC where the helical pitch is much larger than the 
wavelength~\cite{21}. The formula (\ref{eq14}) has been 
generalized~\cite{59,62,85,86} for cholesteric LC 
where the helical pitch is of the same order with the wavelength. The wave 
vector for the Berreman method:
\[
q^\pm =\pm \sqrt {k_\varphi ^2+n^2k_0^2 \pm k_0 \sqrt {4k_\varphi 
^2n^2+\left( {2n\delta n+\delta n^2} \right)^2k_0^2 } } .
\]
It gives the effective Berreman RI:
\begin{equation}
\label{eq30}
n_B ^\pm =\pm \sqrt {n_\varphi ^2+n^2\pm \sqrt {4n_\varphi ^2n^2+\left( 
{2n\delta n+\delta n^2} \right)^2} } .
\end{equation}
The Berreman RI (\ref{eq30}) and the Jones RI (\ref{eq14}) contrast is evident in the graph 
scale despite the reasonable approximation $\varphi \ll\delta \ll\sigma $.

\section{Concluding Remarks}
\label{sec:concluding}
To the best of our knowledge, the Mauguin phase shift formula (\ref{eq13}) has had 
only implicit and indirect confirmations in polarization measurements. Such 
measurements are, for example, the measurement of spectral positions of 
Gooch--Terry minima and the optical activity measurement of cholesteric 
LC~\cite{73}. As usual in commercial TN cells TSS does not exceed 1 nm in 
accordance with Eq. (\ref{eq26}). TSS is really hard to ascertain experimentally. 
For example, in the experimental spectra of TN-FPC with twist-homeotropic 
electric switching, the $o$-polarized spectral shift, as reported 
in~\cite{56}, was much higher than TSS Eq. (\ref{eq26}). The 
crucial factor was reorientation of the parietal LC sublayers. They were not 
reoriented homeotropically up to the cell breakdown voltage.

The reported shift in TN-FPC can be observed directly without any 
polarizers. And the required measurement accuracy is achieved due to 
multiple interference in the cavity. The main drawback of the presented 
scheme is that untwisted structure maintains a constant RI for the 
$o$-polarized light only. The $e$-polarized spectral shift may be measured in the 
experimental scheme with the twist-planar to homogeneous-planar transition, 
which could be achieved using photoalignment material with reversible 
intermolecular bonds~\cite{87,88}.

The avoided crossing spectral shift caused by the mirror reflection mode 
coupling, the above-mentioned RSS, was described analytically and 
experimentally in the uniformly twisted structure. The analytical expression 
(\ref{eq25}) for RSS near the Gooch--Terry maximum looks much easier than the one 
published before~\cite{34}. A generalized Mauguin--Poincar\'{e} rolling 
cone method allowed us to solve the problem geometrically, independently of 
the Jones and Berreman matrix formalisms.

The spectral shift of transmission peaks should not be confused with a 
frequency shift or a frequency conversion. In the stationary linear problem 
the light frequency is not converted. Also we claim that the twist spectral 
shift characterizes not entirely the cavity but namely the twisted layer 
itself. The cavity just facilitates the measurement in that the twisted 
layer does not generate any transmission peaks. So what is actually shifted 
while twisting the layer? Obviously, shifted is the eigenwave phase when 
going out of the twisted layer. Therefore, the effective refractive index 
varies. This optical response can be measured without a cavity. For example, 
a polarization grating has considerable sensitivity to a minute change in 
refractive index, which permits the experimental confirmation of the 
described phenomenon.

The intermediate optical response presumption was formulated to determine 
the extreme values of the optical response of the complex medium. Despite 
the apparent evidence this presumption may be violated. A well-known example 
is a composite of a few optical media with the inhomogeneity scale much 
smaller than the wavelength. Its resonant optical response may exceed 
maximal values for every component. This intermediate optical response 
violation for composites is explained by the Clausius--Mossotti contribution 
of spatial boundaries between the components~\cite{89,90}. Another 
intermediate optical response violation for a twisted anisotropic medium is 
shown to be due to the GP contribution. This tiny GP contribution should not 
be confused with the GP of the zeroth-order adiabatic approximation. The 
last one is responsible for $\pi $-phase polarization conflict in the $\pi 
$-twisted LC cell~\cite{11}. A visual GP representation is speculated 
as the area covered by the polarization trajectory on PS. Contributions of 
TSS and RSS are related to areas of spherical rectangles and triangles. 

The revealed tiny spectral shift in TN-FPC exists in an arbitrary 
anisotropic chiral medium, not only TN layer. No doubt, it must be accounted 
for a vast class of twist-polarization devices in high-precision engineering 
applications, such as multiplexers, 3D~displays, optical tweezers, 
holographic data storage, and diffractive optics.

\section*{Acknowledgements}
This work was supported in part by the Grants of Russian Foundation for 
Basic Research Nos. 14-02-31248 and 15-02-06924; Russian Ministry of 
Education and Science under the Government program, Project No. 
3.1276.2014/K; the Ministry of Science and Technology of Taiwan under Grant 
No. NSC 103-2923-M-009-003-MY3 through an NSC--SB-RAS joint project.

Address correspondence to I. V. Timofeev, Kirensky Institute of Physics, 
Akademgorodok 50/38, Krasnoyarsk, 660036 Russia. E-mail: tiv@iph.krasn.ru

\end{document}